\documentclass[times,review]{elsarticle}

\usepackage{caption}
\usepackage{amssymb}
\usepackage{amsmath}
\usepackage{hyperref}       
\usepackage{url}            
\usepackage{amsfonts}       
\usepackage[margin=1in]{geometry}

\journal{}

\begin{document}

\begin{frontmatter}

\title{ADEPT: A Noninvasive Method for Determining Elastic Parameters of Valve Tissue}

\author[1,2]{Wensi Wu}
\author[5,6]{Mitchell Daneker}
\author[3]{Christian Herz}
\author[3]{Hannah Dewey}
\author[7,8]{Jeffrey A. Weiss}
\author[9,10]{Alison M. Pouch}
\author[5]{Lu Lu\corref{correspondingauthor}}
\author[3,4]{Matthew A. Jolley\corref{correspondingauthor}}

\address[1]{Department of Mechanical Engineering and Applied Mechanics, University of Pennsylvania, Philadelphia, PA, USA}
\address[2]{Cardiovascular Institute, Children's Hospital of Philadelphia, Philadelphia, PA, USA}
\address[3]{Department of Anesthesiology and Critical Care Medicine, Children's Hospital of Philadelphia, Philadelphia, PA, USA}
\address[4]{Division of Cardiology, Children's Hospital of Philadelphia, Philadelphia, PA, USA}
\address[5]{Department of Statistics and Data Science, Yale University, New Haven, CT, USA}
\address[6]{Department of Chemical and Biochemical Engineering, University of Pennsylvania, Philadelphia, PA, USA}
\address[7]{Department of Biomedical Engineering, University of Utah, Salt Lake City, UT, USA}
\address[8]{Scientific Computing and Imaging Institute, University of Utah, Salt Lake City, UT, USA}
\address[9]{Department of Bioengineering, University of Pennsylvania, Philadelphia, PA, USA}
\address[10]{Department of Radiology, University of Pennsylvania, Philadelphia, PA, USA}
\cortext[correspondingauthor]{Corresponding authors. Email: lu.lu@yale.edu, jolleym@chop.edu}

\begin{abstract}
Computer simulation of ``virtual interventions" may inform optimal valve repair for a given patient prior to intervention. However, the paucity of noninvasive methods to determine \textit{in vivo} mechanical parameters of valves limits the accuracy of computer prediction and their clinical application. To address this, we propose ADEPT: \textbf{A} noninvasive method for \textbf{D}etermining \textbf{E}lastic \textbf{P}arameters of valve \textbf{T}issue. In this work, we demonstrated its application to the tricuspid valve of a child. We first tracked valve displacements from open to closed frames within a 3D echocardiogram time sequence using image registration. Physics-informed neural networks were subsequently applied to estimate the nonlinear mechanical properties from first principles and reference displacements. The simulated model using these patient-specific parameters closely aligned with the reference image segmentation, achieving a mean symmetric distance of less than 1 mm. Our approach doubled the accuracy of the simulated model compared to the generic parameters reported in the literature.
\end{abstract}


\end{frontmatter}

\section{Introduction}
Valvular heart disease is a leading contributor to cardiovascular morbidity and mortality, affecting nearly 41 million people worldwide~\cite{Coffey2021, Santangelo2023}, necessitating 20\% of cardiac surgery in the United States. Both heart valve stenosis and regurgitation lead to decreased efficiency of circulation, resulting in heart failure, end-organ (\textit{e.g.,} lung,  kidney) dysfunction, or death~\cite{Tseng2020, arnaert2021}. In this context, valve repair, either surgical or transcatheter, is increasingly applied to restore valvular function ~\cite{Hensey2021, Buratto2021}. However, determining which valves can be repaired and how to best repair an individual valve remains a significant challenge~\cite{Asmarats2018, Del-Forno2020}. Patient-specific image-derived quantitative structural modeling~\cite{Levack2012, Salgo2002, Nam2023, Bouma2016, Wijdh-den-Hamer2016} has improved the understanding of the relationship between 3D valve structure and valve dysfunction in both adults ~\cite{Lee2013, Grewal2010, Levack2012, Salgo2002} and children~\cite{Nam2022ATS, Nam2022JASE, Nguyen2019, Colen2018, Kutty2014} without requiring knowledge of the underlying valve biomechanics. However, these static models cannot alone be used as a predictive tool to realistically explore dynamic valve dysfunction or the result of the application of a potential valve repair technique. In this setting, image-derived, physics-based computational approaches directed toward understanding biomechanical risk factors for valve failure~\cite{narang2021, El-Tallawi2021} and repair optimization have the potential to reduce the need for a trial-and-error approach, and allow identification of the ideal repair for an individual patient before intervention (image-derived precision medicine)~\cite{Biffi2019, Khalighi2019, Kong2018, Kong2020, Sacks2019, Villard2018, Sacks2019}. 

While computational models (\textit{e.g.,} finite element analysis, fluid-structure interactions) elicit the potential for emulation of \textit{in vivo} valve motion and iterative optimization of valve repair\cite{Biffi2019, Khalighi2019, Kong2018, Kong2020, Sacks2019, Villard2018}, the accuracy of image-derived simulations remains contingent upon the mechanical properties of an individual patient's valve tissue, which may vary across age and pathology~\cite{Wu2023_JMBBM}. Unfortunately, existing methods for estimating these properties rely upon destructive \textit{ex vivo} mechanical testing animals or pathological human specimens, which is not clinically possible in individual patients~\cite{Krishnamurthy2008, raush2013, Laville2020, Aggarwal2016, labrosse2016planar, stella2007biaxial}. This methodological gap, in turn, hinders the development of the accurate patient-specific computational models necessary for predicting valve behavior, particularly in children and patients with congenital heart disease where mechanical properties may significantly differ from typical adult populations~\cite{van2016}.

This limitation motivated our development of ADEPT, a generalizable framework that combines image registration~\cite{Yushkevich2016, Aggarwal2023} and physics-informed neural networks (PINNs)~\cite{raissi2019,lu2021,karniadakis2021physics} to noninvasively predict patient-specific tissue parameters from clinically acquired 4D (3D + time) medical images. In this framework, image registration was used to analyze the spatial translation (displacements) of heart valves at different time frames. These image-derived displacements were subsequently supplied to PINNs to predict the underlying elastic parameters of valve tissue. Notably, PINNs synergistically utilize the underlying governing equations (PDEs that define the behavior of the system at work) and empirical observation (experimental data) to optimize the weights and biases within the neural network in the training process and guide the prediction of the material responses and properties~\cite{karniadakis2021, lu2021physics}. The dual regularization from first principles and experimental observations enhances both the accuracy and applicability of the model prediction~\cite{chen2020, yazdani2020systems, lu2021physics, Wu2023_AMM, daneker2023systems, caforio2024, Wu2024_SM, fan2024deep, daneker2024transfer}. As such, PINNs have increasingly been utilized in concert with medical image data to investigate the biophysics and biomechanics of the cardiovascular system~\cite{buoso2021, van2022,liang2023}.

However, to date, such studies have primarily focused on solving forward problems using idealized geometries that do not fully reflect the complexity encountered in clinical contexts. In contrast, our approach uniquely integrates image-derived displacement fields and fundamental principles of continuum solid mechanics within PINNs to estimate patient-specific elastic parameters of complex anatomical structures from clinically acquired datasets. To our knowledge, this work represents the first integration of imaging analysis and PINNs to obtain \textit{in vivo} tissue elastic parameters using clinical images. The proposed approach advances inverse analysis strategies for mechanical properties characterization of biological tissues, extending beyond the ideal conditions afforded by \textit{ex vivo} or \textit{in vitro} experiments and demonstrating promising potential for translational applications. The resulting elastic parameters from ADEPT, in concert with patient-specific valve structure, will facilitate more accurate personalized simulations of \textit{in vivo} heart valve dynamics and systematic comparison of repair strategies in an individual patient~\cite{Wu2022_JBME, Wu2023_JMBBM}. As such, the proposed framework represents a critical translational step toward noninvasive, patient-specific diagnostics and treatment planning, with the potential to benefit both pediatric and adult populations. Notably, while our proof of concept study utilizes 3D echocardiogram (3DE) images of a child with palliated single ventricle heart disease to demonstrate application in a vulnerable population using the most common clinical valve imaging modality (echocardiography), our methodology to determine material parameters is fundamentally applicable to any 4D image of sufficient spatial and temporal resolution. 

This paper is organized as follows. In Section \ref{re:results}, we provide an overview of ADEPT and present the problem descriptions and parameter verification results for each example. In Sections \ref{discussion} and \ref{conclusion}, we discuss the clinical implications, major contributions, limitations and future direction of the current work. In Section \ref{sec:methods}, we provide theoretical background and formulation of image registration, PINNs, and material constitutive models concerning this work.

\section{Methods}
\label{sec:methods}

\subsection{Overview of material parameter identification procedure}
\begin{figure}
\centering
\vspace{-2ex}
\includegraphics[width=1\textwidth]{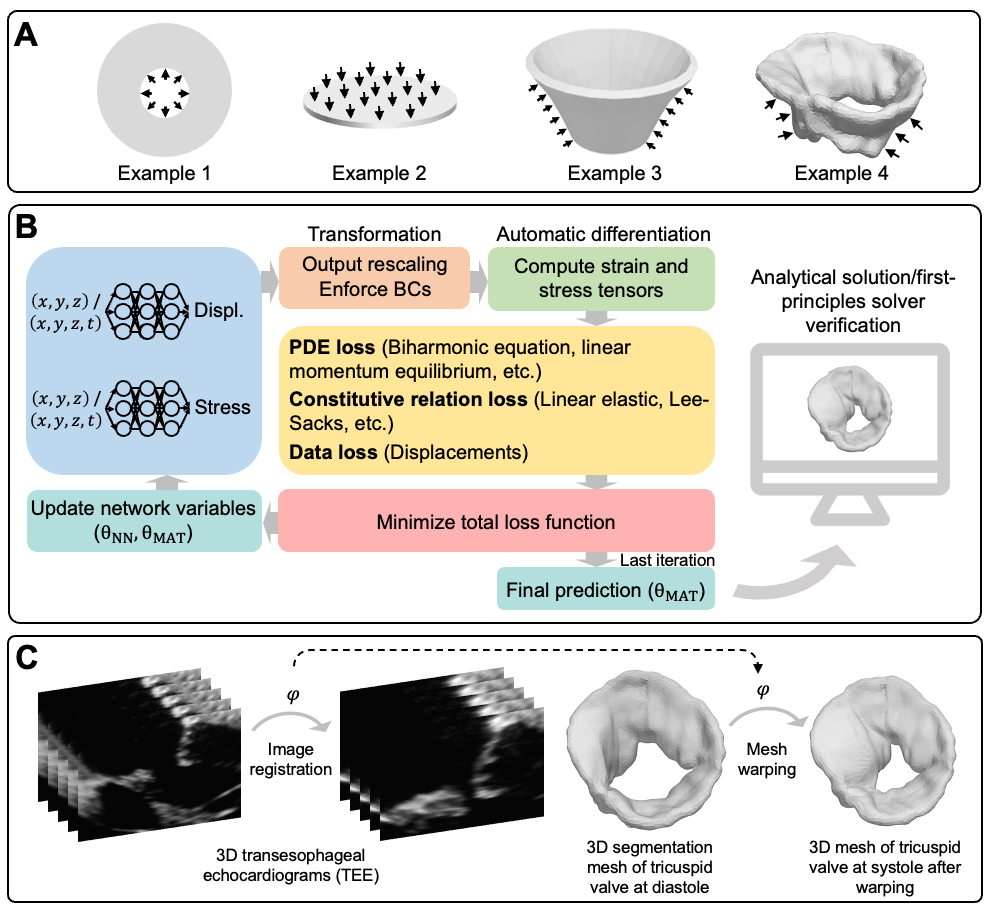}
\vspace{-3ex}
\caption{\textbf{Material parameter identification procedure.} (\textbf{A}) Example cases considered in the current study were a 2D thick-walled cylinder subject to internal pressure (example 1), a 2D thin circular plate subjected to transverse pressure (example 2), a 3D truncated cone subject to external pressure (example 3), and a 3D image-derived patient-specific regurgitant tricuspid valve model subject to transvalvular pressure (example 4). (\textbf{B}) The PINN architecture contained a two-layer parallel network with two feedforward networks that independently predict displacements and stresses. The reference displacement data of the example cases were used to guide the minimization of data loss in PINNs. In examples 1 and 2, the analytical solutions of displacements were sought. Displacements in example 3 were estimated through FEA, and those in example 4 were approximated using deformable image registration illustrated in (\textbf{C}).}\label{fig:overview}
\vspace{-3ex}
\end{figure}

An overview of the proposed method is presented in Fig.~\ref{fig:overview}. Fig.~\ref{fig:overview}A illustrates the geometry of the examples considered in the work; relevant problem formulation and descriptions specific to each example are provided in Sections~\ref{re:2DHollowCylinder} to ~\ref{re:3DTV}. Fig.~\ref{fig:overview}B presents the general workflow of our prediction-verification procedure. In examples 1, 3, and 4, a parallel neural network architecture with two independent feedforward neural networks (FNNs) was chosen for the PINN model, given its efficiency and effectiveness, as demonstrated in our previous work~\cite{Wu2024_SM}. The Cartesian coordinates ($x$, $y$, and $z$) of the object were the input for the FNNs, with one network outputting displacements and the other outputting stresses. We subsequently constructed a loss function by setting up penalty terms on momentum equilibrium, material constitutive balance, and kinematics constraints (\textit{i.e.,} enforcing displacement agreement between predicted values and reference data). At each iteration, the network weights and biases (denoted as $\theta_{\text{NN}}$) and the unknown material parameters (denoted as $\theta_{\text{MAT}}$) are optimized together to reduce the loss. These weights and biases within the networks establish a functional mapping between the coordinates and the resultant displacements and stresses. In example 2, only the displacement network was needed to inform prediction. As such, the material constitutive balance was ignored in this example. 

We computed the reference displacements of the first and second examples from analytical formulations. Displacements from the third example were computed using finite element analysis (FEA), whereas those of patient-specific tricuspid valves were estimated using an image registration approach, described in Fig.~\ref{fig:overview}D. In particular, 3D TEE image volumes of two consecutive frames were registered using a non-symmetric diffeomorphic image registration algorithm~\cite{Yushkevich2016}. This step resulted in a voxel-wise transformation matrix, $\varphi$, that maps the image intensity from the moving to the fixed 3D TEE image volume. The transformation matrix, $\varphi$, was subsequently applied to a manual segmentation of the open leaflets in order to estimate the closed leaflet geometry. The differences between the nodal points of the open and closed valves were calculated, serving as reference displacement fields.  

\subsection{Image processing}
Here, we outline the image analysis process to acquire patient-specific geometry and displacement fields of the tricuspid valve. All image processing tasks were performed using open-source image analysis platforms, 3D Slicer~\cite{Fedorov2012}, SlicerHeart~\cite{Lasso2022}, and Greedy deformable image registration~\cite{Yushkevich2016}.

\subsubsection{Image acquision}
The 3D transesophageal (TEE) images of a regurgitant tricuspid valve were identified from an existing database at the Children's Hospital of Philadelphia. The images were acquired on a Philips Epiq system (Philips Medical, Boston, MA) from a child with Hypoplastic left heart syndrome (HLHS). HLHS is a complex congenital heart disease that affects more than 1,000 liveborns in the US each year. Patients with HLHS have to undergo three staged open-chest reconstruction surgeries. About twenty-five percent of HLHS patients develop tricuspid regurgitation following the third surgery, which significantly increases their risk of heart failure. The objective of this work is to provide a noninvasive approach to evaluate the mechanical factors that influence valve competency in HLHS patients. This study was approved by the Institutional Review Board at the Children's Hospital of Philadelphia. The 3D TEE images were imported into 3D Slicer using the Philips 4D US DICOM patcher module in SlicerHeart~\cite{Scanlan2018, Nguyen2019} for image segmentation and registration.

\subsection{Manual image segmentation}
The selected 3DE images were manually segmented using the SliceHeart extension 3D Slicer. Specifically, an expert observer manually traced the leaflets of the tricuspid valve in the last diastolic frame of the cardiac cycle. The image voxels that correspond to the leaflets were assigned a label value of 1, and all other voxels had a label of 0. The visible segments (\textit{i.e.,} image voxels with label 1) were converted to a surface model that encapsulates the full 3D geometry of the tricuspid valve. The segmentation model was subsequently meshed with 5000 nodes using the Surface Toolbox module in 3D Slicer to create a smooth representation of the tricuspid valve geometry.

\subsection{Deformable image registration}
Intensity-based deformable registration was performed on two consecutive cardiac frames (the last open frame and the first closed frame were chosen for this study) to obtain the deformation fields of the tricuspid valves when fully pressurized. The Greedy diffeomorphic image registration algorithm was used to facilitate deformable registration between the 3DE reference-target image pair~\cite{Yushkevich2016}. The image deformation approach performed in this work was inspired by Aggarwal et al.~\cite{Aggarwal2023}. In particular, we first defined the registration region of interest (ROI) for the reference frame by dilating the diastolic segmentation of the tricuspid valve with 10 voxels in the $x$, $y$, and $z$ directions. To define the registration ROI for the target frame, we roughly approximated the deformation fields by registering the 3DE reference grayscale image at half-resolution and warped the resulting deformation fields to the reference ROI. Finally, we performed full-resolution registration on the grayscale image pair, guided by their respective ROI, to acquire refined deformation fields on the tricuspid valve leaflets. 

The Greedy registration framework is inspired by the Large Deformation Diffeomorphic Metric Mapping (LDDMM) formulation presented in~\cite{Avants2011}. For a given image pair, suppose that the reference image is denoted $\mathcal{I}$, the target image is denoted $\mathcal{J}$, and a diffeomorphism map that transforms the physical coordinates of image $\mathcal{I}$ into image $\mathcal{J}$ is defined as $\varphi$. The diffeomorphism, $\varphi$, is obtained by integrating the velocity fields, $\mathbf{v_t}$, from time $t_0$ to $t_1$ defined as 
\begin{equation*}
\varphi_{t_1} = \varphi_{t_0} + \int_{t_0}^{t_1}\|\mathbf{v}_t\|_L dt,
\end{equation*}
where $\|\cdot\|_L$ is a linear differential operator that regularizes the velocity field, expressed in the form of $L = \alpha\nabla^2+\beta\textbf{Id}$. Within linear operator $L$, \textbf{Id} is the identity map. $\alpha$ and $\beta$ are gradient smoothing and deformation field smoothing constants, with assigned values of 2.3mm and 0.35mm, respectively, to ensure good registration alignment while preserving smooth displacement features. The objective is to compute velocity fields, $\mathbf{v_t}$, such that the image similarly loss, $\mathcal{L}_\text{sim}(\varphi \mathcal{I} - \mathcal{J}$), is minimized. The sum of the squared difference similarity measure was used in the present work. The Greedy registration tool shares many similarities with the theoretical framework established in~\cite{Avants2008}, while also incorporating additional image metrics and nonsymmetric deformation implementations to optimize computational efficiency~\cite{joshi2004}.

\subsection{Physics-informed neural networks for material parameter identification} \label{pinns_formulation}
Here, we provide an overview of PINNs and the technical details considered in this work. The PINN architecture was set up using the DeepXDE library~\cite{lu2021}. All PINN experiments were trained on an NVIDIA H100 80GB GPU. The source code will be made available upon publication in the GitHub repository \url{https://github.com/lu-group/adept}.

\subsubsection{Neural network architecture}
The PINN architecture is shown in Fig.~\ref{fig:overview}B. Let $\mathcal{N}^{L}(\mathbf{x}) \colon \mathbb{R}^{\dim(\mathbf{x})} \to \mathbb{R}^{\dim(\mathbf{y})}$ be a $L$-layer neural network that maps input features $\mathbf{x}$ to output feature $\mathbf{y}$ with $\mathcal{N}^l$ neurons in the $l$-layer. The connectivity between layer $l$ and $l-1$ is governed by $\mathcal{N}^l(\mathbf{x}) = \phi(\mathbf{W}^l\mathcal{N}^{l-1}(\mathbf{x})+\mathbf{b}^l)$, where $\phi$ is a nonlinear activation function, $\mathbf{W}^{l}$ is a weight matrix, and $\mathbf{b}^l$ is a bias vector. We used \textit{swish} activation function and Glorot uniform weight initialization method in all analyses. Given that the activation function is applied element-wise to each neuron, the recursive FNN is defined as:
\begin{align*}
\textrm{input layer}: \quad & \mathcal{N}^0(\mathbf{x}) = \mathbf{x} \in \mathbb{R}^{\dim{\mathbf{(x)}}}, \\
\textrm{hidden layer $l$}: \quad & \mathcal{N}^l(\mathbf{x}) = \text{swish} \left(\mathbf{W}^l\mathcal{N}^{l-1}(\mathbf{x})+\mathbf{b}^l \right) \in \mathbb{R}^{\mathcal{N}^l}, \quad \textrm {for} \quad 1\leq l \leq L-1, \\
\textrm{output layer}: \quad & \mathcal{N}^L(\mathbf{x}) = \mathbf{W}^L\mathcal{N}^{L-1}(\mathbf{x})+\mathbf{b}^L \in \mathbb{R}^{\dim{(\mathbf{y})}}.
\end{align*}
The nodal coordinates were used as input variables for the network. The network architectures consisted of two independent feedforward networks. One of the independent networks was responsible for estimating displacement fields $\mathcal{N}_{u_i}$. The other independent network was responsible for estimating the stress fields $\mathcal{N}_{\sigma_{ij}}$. The displacement fields were normalized based on the mean and standard deviation of the ground truth displacement data supplied to improve training efficiency. Within the architecture, $\theta_\text{NN}$ encapsulates network variables $\mathbf{W}^{l}$ and $\mathbf{b}^l$, while $\theta_\text{mat}$ contains unknown material variables.

\subsubsection{Loss function}

In inverse analysis, PINN seeks to optimize the network parameters, $\theta_{\text{NN}}$ (\textit{i.e.,} $\mathbf{W}^l$ and $\mathbf{b}^l$), and the unknown material parameters, $\theta_{\text{mat}}$ (\textit{e.g.,} $E$ and $\nu$) in the training process expressed as
\begin{equation*}
\theta_{\text{NN}}^*, \theta_{\text{mat}}^* = \underset{\theta_{\text{NN}}, \theta_{\text{mat}}}{\arg\min} \mathcal{L}(\theta_{\text{NN}}, \theta_{\text{mat}}),
\end{equation*}
with the general total loss function $\mathcal{L}(\theta_{\text{NN}}, \theta_{\text{mat}})$ defined as 
\begin{equation*}
\mathcal{L}(\theta_{\text{NN}}, \theta_{\text{mat}}) =
w_{\text{PDEs}}\mathcal{L}_{\text{PDEs}} +
w_{\text{M}}\mathcal{L}_{\text{M}} +
w_{\text{F}}\mathcal{L}_{\text{F}} +
w_{\text{D}}\mathcal{L}_{\text{D}}.
\end{equation*}
Herein, $w_\bullet$ represents the weight associated with its corresponding loss term $\mathcal{L}_\bullet$. $\mathcal{L}_{\text{M}}$, $\mathcal{L}_{\text{F}}$, and $\mathcal{L}_{\text{D}}$ refer to errors in material constitutive relations, traction balance, and displacement reference data, respectively. The strong form of the momentum equation was selected as the governing PDEs. The Dirichlet conditions were implicitly enforced in the PDE approximation as hard constraint conditions. Note that the traction loss was only considered in examples 1 and 2. Examples 3 and 4 were trained without enforcing force balance to evaluate the robustness of our approach in settings where the underlying physics was only partially known. 
 
In examples 1--3, all loss terms were optimized using mean squared errors. In example 4, the loss term $\mathcal{L}_{\text{D}}$ was optimized by MSD as image intensity-derived displacement fields may differ from the displacement fields derived from material point frame of reference in continuum mechanics~\cite{Rego2022}. Suppose we have two finite point sets, $P$ and $R$, the MSD is defined as
\begin{equation*}
\mathcal{L}_{\text{D}} = \frac{1}{2}\left(\frac{1}{|P|}\sum_{p \in P}\min_{r \in R}\|p - r\| + \frac{1}{|R|}\sum_{r \in R}\min_{p \in P}\|p - r\|\right).
\end{equation*}
The MSD measures the average minimum distance from $P$ to $R$ and from $R$ to $P$. Here, $\|\cdot\|$ denotes the Euclidean distance. The point sets $P$ and $R$ represent the deformed nodal coordinates of the tricuspid valve at systole estimated in PINNs and derived from image registration, respectively. 

\subsection{Material constitutive models}
Various material constitutive models were tested in the examples presented in this work. The benchmark examples detailed in Sections~\ref{re:2DHollowCylinder} to \ref{re:3DTrucatedCone} assumed an isotropic linear elastic material model. For the patient-specific tricuspid valve described in Section~\ref{re:3DTV}, isotropic Neo-Hookean and Lee-Sacks material models were applied.

\paragraph{Linear elastic model.} The isotropic linear elastic material constitutive model is defined as:
\begin{equation*}
\mathbf{\sigma} = \mathbf{C} \cdot \mathbf{\epsilon},
\end{equation*}
with 
\begin{gather*}
\mathbf{\sigma} = 
\begin{bmatrix}
\sigma_{xx} \\ \sigma_{yy} \\ \sigma_{xy}
\end{bmatrix}; \quad 
\mathbf{C} = \frac{E}{(1-\nu^2)}
\begin{bmatrix}
1 & \nu & 0 \\ 
\nu & 1 & 0 \\ 
0 & 0 & (1-\nu)
\end{bmatrix};
\quad
\mathbf{\epsilon} = 
\begin{bmatrix}
\epsilon_{xx} \\ \epsilon_{yy} \\ \epsilon_{xy}
\end{bmatrix}.
\end{gather*}
In the 2D thick-walled cylinder and 3D truncated cone examples, the kinematic relations are expressed as:
\begin{gather*}
\epsilon_{xx} = \frac{\partial u_x}{\partial x}; \quad \epsilon_{yy} = \frac{\partial u_y}{\partial y}; \quad
\epsilon_{xy} = \frac{1}{2}[\frac{\partial u_x}{\partial y}+\frac{\partial u_y}{\partial x}].
\end{gather*} 
In the deflected circular plate example, $u_x = -z\frac{\partial u_z}{\partial x}$ and $u_y = -z\frac{\partial u_z}{\partial y}$.  Therefore, the kinematic relations are
\begin{gather*}
\epsilon_{xx} = -z\frac{\partial^2 u_z}{\partial x^2}; \quad \epsilon_{yy} = -z\frac{\partial^2 u_z}{\partial y^2}; \quad
\epsilon_{xy} = -z\frac{\partial^2 u_z}{\partial x \partial y}.
\end{gather*} 

\paragraph{Neo-Hookean model.} The Neo-Hookean model~\cite{bonet_wood_2008} is developed to characterize the behavior of nonlinear materials that undergo deformations. Within this model, the strain energy density function, $\Psi$, is defined as
\begin{equation*} 
\Psi(I_1, J) = \frac{1}{2}\lambda[log(J)]^2-\mu log(J)+ \frac{1}{2}\mu (I_1-3),
\end{equation*}
where $I_1$ is the first principal invariants denoted as $I_1 = \text{trace}(\mathbf{F}^T\cdot \mathbf{F})$, $\mathbf{F}$ is the deformation gradient denoted as $F_{ij} = \delta_{ij} + u_{i, j}$, $u_i$ is the displacement vector, and $\lambda$ and $\mu$ are the Lam\'e's elasticity parameters. The first Piola-Kirchhoff stress is given by
\begin{equation*}
P = \frac{\partial \Psi}{\partial \mathbf{F}} = \mu \mathbf{F}+[\lambda log(J)-\mu]\mathbf{F}^{-T}.
\end{equation*}
where
\begin{gather*}
\lambda = \frac{E\nu}{(1+\nu)(1-2\nu)}, \quad \textrm{and} \quad \mu = \frac{E}{2(1+\nu)}. 
\end{gather*}

\paragraph{Lee-Sacks model.} The Lee-Sacks isotropic material model~\cite{lee2014} is a popular model for characterizing heart valve tissue properties. It combines Neo-Hookean with an exponential term to account for contributions of the extracellular matrix and collagen fiber network in leaflet tissues. The hyperelastic strain energy function is expressed as
\begin{equation*}
\Psi(I_1)= \frac{c_0}{2}(I_1-3)+\frac{c_1}{2}\{\exp{[c_2(I_1-3)^2]}-1\},
\end{equation*}
where $c_0$, $c_1$, and $c_2$ are stiffness parameters. The first Piola-Kirchhoff stress is given by
\begin{equation*}
P =  \frac{\delta \Psi}{\delta \mathbf{F}} = (c_0+2c_1c_2(I_1-3)\exp{[c_2(I_1-3)^2]})\mathbf{F}.
\end{equation*}

\section{Results}
\label{re:results}
We considered four examples in this work: a 2D thick-walled cylinder subjected to internal pressure, a 2D thin circular plate subjected to transverse pressure, a 3D truncated cone subjected to external pressure, and a 3DE-derived tricuspid valve subjected to transvalvular pressure. The first three benchmarks were selected to systematically increase the complexity presented by the final patient-specific tricuspid valve example. Examples 1 and 2 were selected to assess the accuracy of solutions for 2D geometries subjected to applied pressures, mimicking the loading conditions commonly observed in heart valves. Example 3 further increases geometric complexity by transitioning from a 2D shape to a 3D geometry, representing an idealized tricuspid valve structure.

In each example, we trained a PINN model to determine the ``unknown" elastic properties from reference displacement data. To verify solution accuracy and improve the interpretability of PINN predictions, we computed the mechanical responses, specifically displacements and stresses, through analytical formulation (if it exists) or first-principle solvers (\textit{e.g.,} FEBio finite element package~\cite{Maas2012}) using the estimated material parameters. The mechanical responses were subsequently compared to the reference solution to evaluate their agreement. All PINN experiments were trained on an NVIDIA H100 80GB GPU.

\subsection{2D thick-walled cylinder subjected to internal pressure}
\label{re:2DHollowCylinder}
\subsubsection{Problem description}
\begin{figure}[!ht]
\centering
\includegraphics[width=0.9\textwidth]{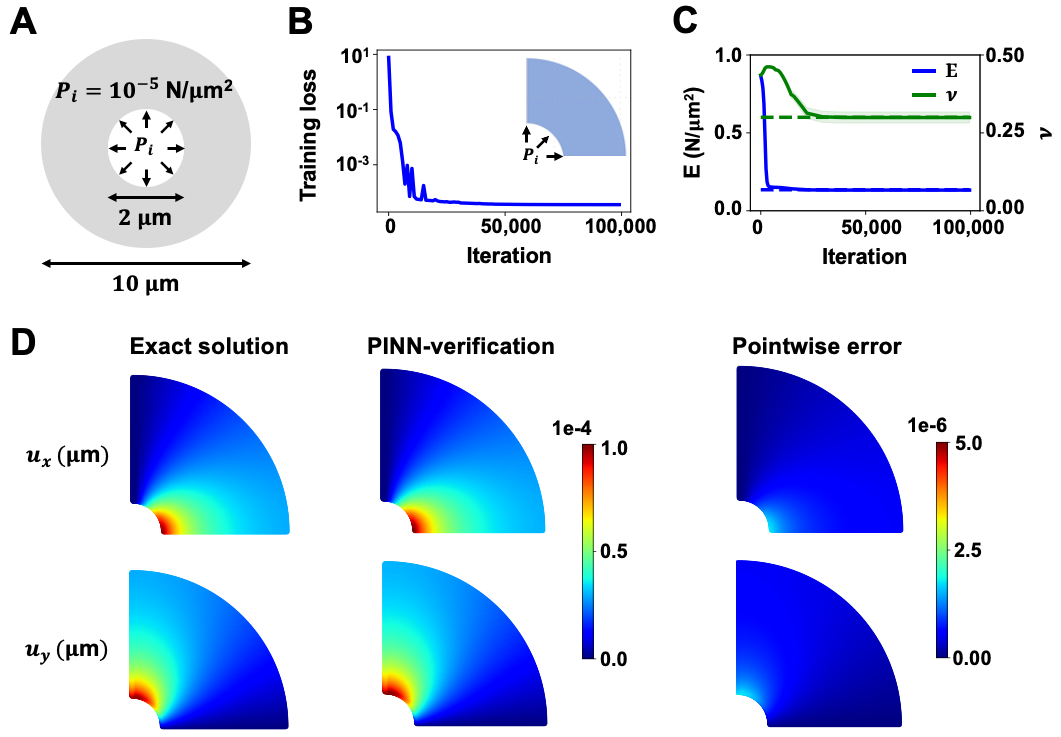}
\caption{\textbf{2D thick-walled cylinder verification results.} (\textbf{A}) A hollow cylinder with 1~$\mu \text{m}$ inner radius and 5~$\mu\text{m}$ outer radius is shown. A uniform pressure load of $10^{-5}\text{N/}\mu\text{m}^2$ is applied to the wall of the inner radius of the cylinder. (\textbf{B}) The training loss history is shown. The total training loss converges to below $10^{-3}$ at 100000 iterations. (\textbf{C}) The average and standard deviation of $E$ and $\nu$ estimated from 10 repeated PINN analyses (solid lines) are plotted along with the ground truth values (dotted lines). The relative errors of $E$ and $\nu$ were 1.72$\%$ and 0.23$\%$, respectively. (\textbf{D}) The exact solution, PINN-verification, and pointwise errors of displacement are presented. The estimated $E$ and $\nu$ produced highly accurate displacementfields with pointwise errors in $\mathcal{O}(10^{-6})$.}\label{fig:example1}
\end{figure}

We started with a 2D linear elastic example, as illustrated in Fig.~\ref{fig:example1}A, to verify the efficacy of our proposed methods. The strong-form static momentum equation, $\sigma_{ij,j}=0$, was the governing PDE, where $\sigma_{ij}$ represents the Cauchy-stress. A thick-walled cylinder with an inner radius of 1 $\mu \text{m}$ and an outer radius of 5 $\mu \text{m}$ was subjected to $10^{-5}$ $\text{N/}\mu\text{m}^2$ uniform internal pressure. The ground truth solution of the thick-walled cylinder was calculated with Young's modulus and Poisson's ratio of 0.135 $N/\mu \text{m}^2$ and 0.3, respectively. Here, we reduced our analysis to one-quarter of the cylinder due to symmetry in geometry and pressure load. An isotropic linear elastic material model was assumed. 

\subsubsection{Estimated results from PINNs}
We adopted a parallel architecture with two independent neural networks. Each independent network within PINN was assigned 5 hidden layers with 45 neurons per layer. The loss weights were $w_\text{PDEs} = 1$, $w_\text{M} = 10$, and $w_\text{D} = 1$. The subscript $M$ refers to material model and $D$ refers to displacement reference data. The learning rate decay was set to $10^{-3}$ with a decay rate of 0.15 every 15,000 iterations. The neural networks were trained using the Adam optimizer for 100,000 iterations. The total training loss, which includes losses in the PDEs, stress-strain relations, reference data observation, and traction condition, converged to a satisfactory value of below $10^{-3}$ at the end of the analysis, as demonstrated in Fig.~\ref{fig:example1}B. The PINN inverse analysis was repeated 10 times with varying random seeds to evaluate its robustness and precision in different network weight initialization. A total of 1,500 PDE and 1,500 data collocation points were used in the training. The average runtime of the 10 repeated tests was 0.5 hours.

The prediction histories of the mean and standard deviation of Young's modulus ($E$) and Poisson's ratio ($\nu$) are presented in Fig.~\ref{fig:example1}C. The dotted lines denote the ground truth values, while the solid lines denote PINN estimation. Similar to our previous work~\cite{Wu2023_AMM, Wu2024_SM}, we constrained the search range of $E$ and $\nu$ to a reasonable range of [0, 1]~$\text{N/}\mu\text{m}^2$ and [0, 0.5], respectively, to facilitate the optimization effort. The mean and standard deviation values of $E$ and $\nu$ of the 10 repeated tests were 0.133$\pm$0.002~$\text{N/}\mu\text{m}^2$ and 0.299$\pm$0.018, respectively. The relative error of the average $E$ was 1.72$\%$, and the average $\nu$ was 0.23$\%$.

\subsubsection{Verification against analytical solution}
Fig.~\ref{fig:example1}D provides the displacement and stress responses between the ground truth and the verification results with PINN-estimated $E$ and $\nu$. The analytical solutions for displacements in Cartesian coordinates were calculated from the following forms: 
\begin{align*}
u_x(x, y) &= \frac{r_i^2P_ir}{E(r_o^2-r_i^2)}[1-\nu + (\frac{r_o}{r})^2(1+\nu)]\text{cos}(\theta), \\
\text{and} \indent 
u_y(x, y) &= \frac{r_i^2P_ir}{E(r_o^2-r_i^2)}[1-\nu + (\frac{r_o}{r})^2(1+\nu)]\text{sin}(\theta).
\end{align*}
In this paper, $r_i$ and $r_o$ are the inner and outer radius. $r$ and $\theta$ are the equivalent radius and angle in polar coordinates, defined as $r=\sqrt{x^2+y^2}$ and $\theta = \text{arctan}(y/x)$. $P_i$ is the internal pressure. The estimated displacements using material parameters derived from PINNs were in excellent agreement with the analytical solution. The maximum pointwise errors of displacements were $\mathcal{O}(10^{-6})$, and the relative errors $L^2$ of $u_x$ and $u_y$ were within 1.71$\%$. 

\subsection{Deflected circular plate subjected to uniform pressure}
\label{re:DeflectedPlate}
\subsubsection{Problem description}
\begin{figure}[!ht]
\centering
\includegraphics[width=0.9\textwidth]{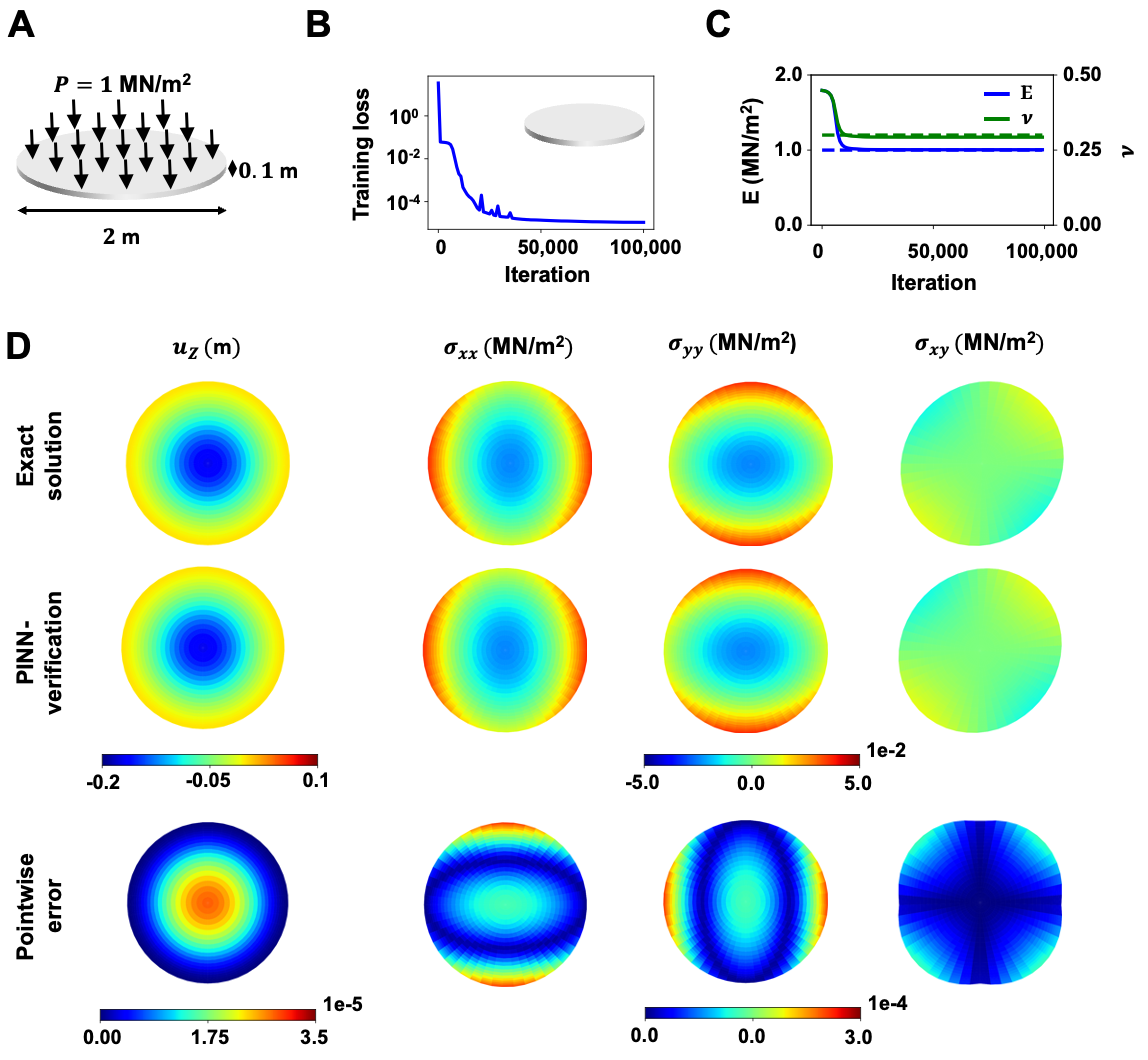}
\caption{\textbf{Deflected circular plate verification results.} (\textbf{A}) A thin circular plate with 1~$\text{m}$ radius and 0.1~$\text{m}$ thickness is shown. A uniform pressure load of $1\text{MN/m}^2$ is applied to its top surface. (\textbf{B}) The training loss history is shown. The total training loss converges to below $10^{-4}$ at the end of analysis. (\textbf{C}) The average and standard deviation of $E$ and $\nu$ estimated from 10 repeated PINN analyses (solid lines) are plotted along with the ground truth values (dotted lines). The relative errors of $E_{\text{mean}}$ and $\nu_{\text{mean}}$ were 0.312$\%$ and 1.489$\%$, respectively. (\textbf{D}) The exact solution, PINN-verification, and pointwise errors of displacement and stress are presented. The absolute pointwise error of the out-of-plane displacements and in-plane stresses are in $\mathcal{O}(10^{-5})$ and $\mathcal{O}(10^{-4})$, respectively.}\label{fig:example2}
\end{figure}

The second benchmark considered in this work was a thin circular plate with 1~$\text{MN/m}^2$ uniform pressure applied to its top surface, illustrated in Fig.~\ref{fig:example2}A. The biharmonic equation, $\nabla^2 \nabla^2 u_z=\frac{q}{D}$ was taken as the governing PDEs of this system. Herein, $u_z$ is the out-of-plane displacement, $q$ is the uniform applied pressure, $\sigma_{ij}$ is the Cauchy-stress, and $D$ is the plate flexural rigidity. Further, $D$ is given by $D = \frac{H^3E}{12(1-\nu^2)}$ with $H$ denotes plate thickness. The circular plate has a radius of 1 $\text{m}$ and a thickness of 0.1 $\text{m}$. It was subjected to a clamped boundary condition along its edge, and an isotropic linear elastic material model was assumed. The ground truth out-of-plane displacements of the circular plate were determined using $E$ and $\nu$ of 1 $\text{MN/m}^2$ and 0.3, respectively.

\subsubsection{Estimated results from PINNs}
In this example, given the direct relation between the out-of-plane displacements and the elastic parameters in the biharmonic equation, only a displacement network was needed to facilitate the parameter prediction. As such, we only considered the PDE loss in the biharmonic equation and data loss. The loss weights were $w_\text{PDEs} = 1$, and $w_\text{d} = 10^3$. The learning rate decay was set to $10^{-3}$ with a decay rate of 0.15 every 15,000 iterations. The neural networks were trained using the Adam optimizer for 100,000 iterations. The total training loss converged to a satisfactory value of below $10^{-4}$ at the end of the analysis, as demonstrated in Fig.~\ref{fig:example2}B. A total of 5,000 PDE and 5,000 data collocation points were used in the training. The average runtime for this benchmark was 1.75 hours.

The prediction histories of the mean and standard deviation of $E$ and $\nu$ are presented in Fig.~\ref{fig:example2}C. We defined a reasonable search range of [0, 2]~$\text{N/}\mu\text{m}^2$ and [0, 0.5] for $E$ and $\nu$, respectively. The mean and standard deviation values of $E$ and $\nu$ of the 10 repeated tests were 1.003$\pm$0.002~$\text{N/}\mu\text{m}^2$ and 0.296$\pm$0.002, respectively. The relative error of $E_{\text{mean}}$ was 0.311$\%$, and $\nu_{\text{mean}}$ was 1.489$\%$.

\subsubsection{Verification against analytical solution}
Fig.~\ref{fig:example2}D provides the out-of-plane displacement and the in-plane stress responses on the top surface of the plate (stresses of the mid-surface are zero based on the Kirchhoff-Love plate theory) between the ground truth and the verification results. The out-of-plane displacements were calculated from the following analytical forms: 
\begin{equation*}
u_z(x, y) = \frac{q}{64D}[a^2 - (x^2+y^2)]^2,
\end{equation*}
where $a$ is the radius of the plate. The analytical solutions of in-plane stresses on the top surface are given by: 
\begin{align*}
\sigma_{xx}(x, y, z) &= \frac{-Ez}{1-\nu^2}(\frac{\delta^2u_z}{\delta x^2}+ \nu \frac{\delta^2u_z}{\delta y^2}), \\
\sigma_{yy}(x, y, z) &= \frac{-Ez}{1-\nu^2}(\frac{\delta^2u_z}{\delta y^2}+ \nu \frac{\delta^2u_z}{\delta x^2}), \\
\text{and} \indent 
\sigma_{xy}(x, y, z) &= \frac{-Ez}{1+\nu}\frac{\delta^2u_z}{\delta x \delta y},
\end{align*}
with $z = 0.05$, half the plate thickness. The $L^2$ relative error of $u_z$ was 0.019$\%$. The $L^2$ relative errors of $\sigma_{xx}$, $\sigma_{yy}$, $\sigma_{xy}$ were 0.378$\%$, 0.387$\%$, and 0.638$\%$, respectively.

\subsection{3D truncated cone subjected to external pressure}
\label{re:3DTrucatedCone}
\subsubsection{Problem description}
\label{re:3DTrucatedCone:pd}
\begin{figure}[!ht]
\centering
\includegraphics[width=0.9\textwidth]{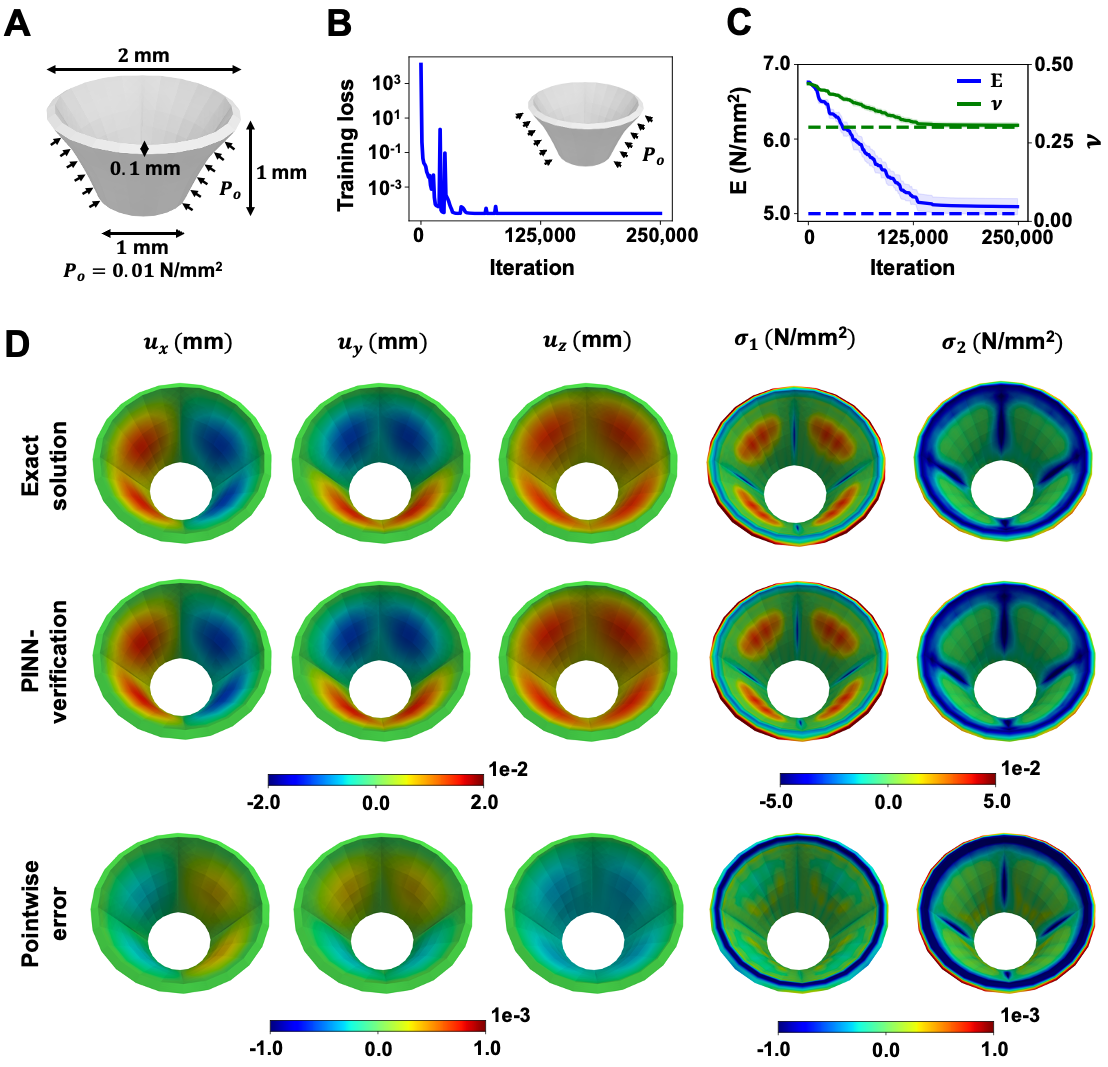}
\caption{\textbf{3D truncated cone verification results.} (\textbf{A}) A hollow 3D truncated cone is shown. The cone is 1~mm tall and 0.1~mm thick. The diameters of the top and bottom end are 2~mm and 1~mm, respectively. A uniform external pressure of $10^{-2} \text{N/mm}^2$ is applied to the outer surface of the cone, with pinned boundary conditions assumed at the top edge of the cone. (\textbf{B}) The total training loss converges rapidly from $\mathcal{O}(10^{3})$ to $\mathcal{O}(10^{-3})$ at the end of the analysis. (\textbf{C}) The relative errors of the PINN-estimated $E$ and $\nu$ (solid lines) were found to be 1.878$\%$ and 2.052$\%$, respectively. (\textbf{D}) The estimated $E$ and $\nu$ produced highly accurate displacement and stress fields with less than 3.5$\%$ $L^2$ relative errors between the displacement components and less than 1.5$\%$ $L^2$ relative errors across the principal stress components.}\label{fig:example3}
\end{figure}

As a step towards a more realistic representation of the tricuspid valve geometry, we modified our PINN model to estimate the material parameters of a hollow 3D truncated cone, displayed in Fig.~\ref{fig:example3}A. Similarly, the strong-form static momentum equation, $\sigma_{ij,j}=0$, was selected as the governing PDE due to quasi-static loading and small deformation nature of this example. The 3D truncated cone was 1~mm tall with a 0.1~mm thick outer surface. The diameters of the top and bottom ends were 2~mm and 1~mm, respectively. The ground truth solution was generated with Young's modulus and Poisson's ratio of 5~$\text{N/mm}^2$ and 0.3, respectively. Pinned boundary conditions ($u_x$=$u_y$=$u_z$=0) were prescribed to the top edge of the cone. A uniform external pressure of 0.01~$\text{N/mm}^2$ was applied to the external surface of the cone. An isotropic linear elastic material model was assumed. 

\subsubsection{Estimated results from PINNs}
A parallel architecture with two independent neural networks was adopted. Each independent network in PINN was made of three hidden layers containing 32, 16, and 8 neurons, respectively~\cite{caforio2024}. The loss weights were $w_\text{PDEs} = w_\text{M} = 10^{-4}$ and $w_\text{D} = 1$. The learning rate decay was set to $10^{-3}$ with a decay rate of 0.66 every 5,000 iterations. The neural networks were trained using the Adam optimizer for 250,000 iterations. The total training loss decreased from $\mathcal{O}(10^3)$ to $\mathcal{O}(10^{-3})$, as shown in Fig.~\ref{fig:example3}B. 

Unlike the previous example, in which the pressure condition was included in the PINN training, we omitted the pressure information from the training process in this example to evaluate the capability of estimating the material parameters of PINNs with limited information when the force is unknown. While the relative errors of the estimated values were higher in this example, due to both the increased complexity in the model geometry and limited information on the applied load, the parameters of the material estimated by PINN were still in excellent agreement with the ground truth, as shown in Fig.~\ref{fig:example2}C. The search range of $E$ and $\nu$ were constrained to the range of [1.5, 7.5]~$\text{N/mm}^2$ and [0, 0.5], respectively. The mean and standard deviation values of $E$ and $\nu$ from 10 repeated tests were 5.093$\pm$0.107~$\text{N/mm}^2$ and 0.306$\pm$0.008, respectively. The average runtime in this example was 1.18 hours. The relative errors of the average $E$ and $\nu$ were 1.878$\%$ and 2.052$\%$, respectively. 

\subsubsection{Finite element verification}
The displacements and principal stresses between the ground truth and verification results are shown in Fig.~\ref{fig:example3}D. These results were obtained using FEA, where the exact solution was generated using the pre-defined material parameters stated in Section.~\ref{re:3DTrucatedCone:pd}, and the PINN-verification was simulated using the PINN-estimated parameters indicated in Fig.~\ref{fig:example3}C. The finite element model of the cone was discretized in 11,520 hexahedron elements and analyzed through quasi-static analysis.

As shown in Fig.~\ref{fig:example3}D, the estimated displacements and principal stresses demonstrate strong alignment with the exact solution, with pointwise errors within the range of $\mathcal{O}(10^{-3})$. The $L^2$ relative errors for $u_x$, $u_y$, and $u_z$ were found to be 2.306$\%$, 2.304$\%$, and 3.103$\%$, respectively; the verification results had lower displacement extrema compared to the exact solution. In addition, the $L^2$ relative error for the first principal stress ($\sigma_1$) was 1.022$\%$, and for the second principal stress ($\sigma_2$) was 0.746$\%$. The highest errors in stresses were observed at the top edge where the pinned boundary conditions were applied. 

\subsection{3D Image-derived tricuspid valve subjected to ventricular pressure}
\label{re:3DTV}
\subsubsection{Problem description}
\begin{figure}
\centering
\includegraphics[width=1\textwidth]{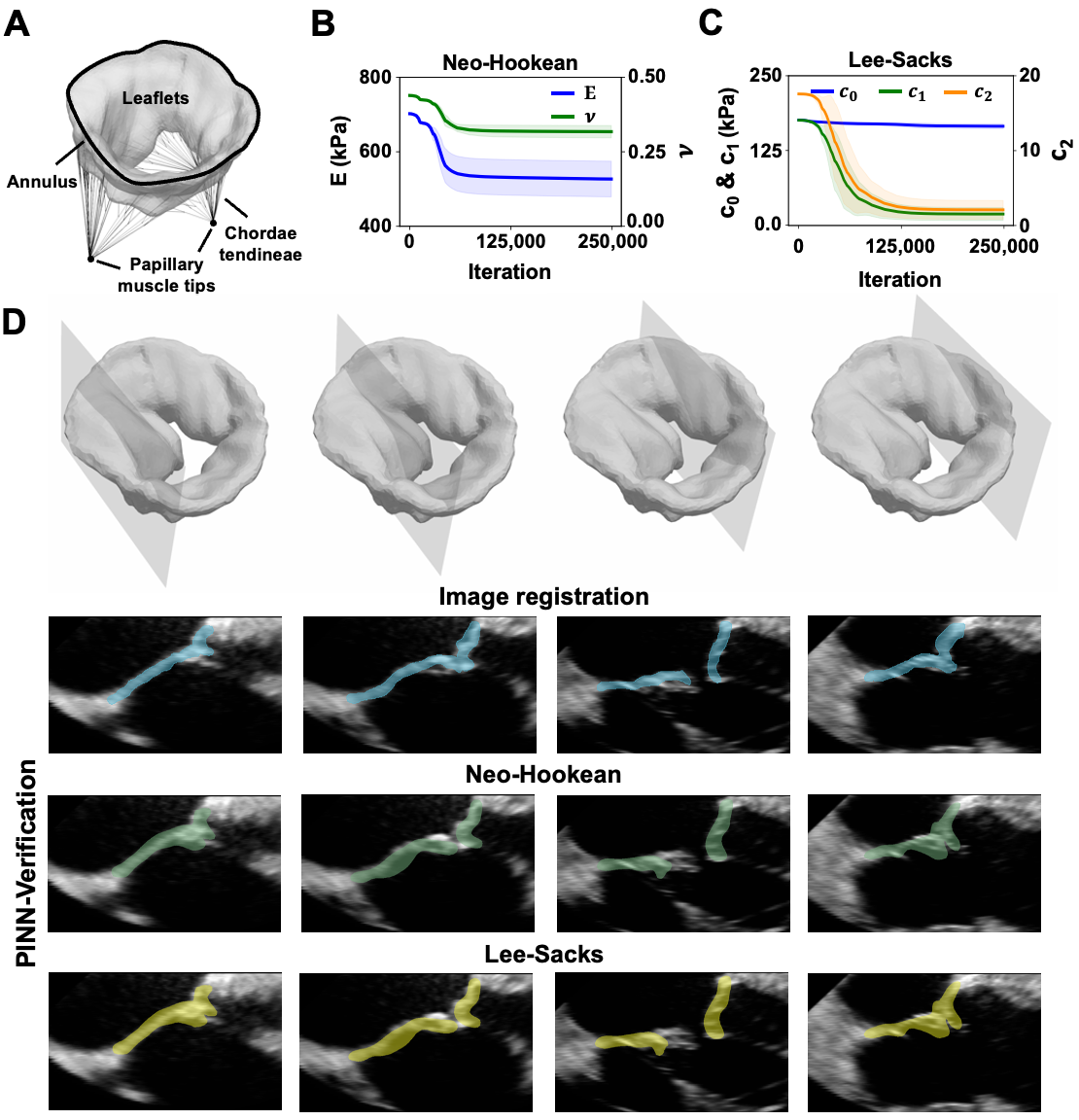}
\caption{\textbf{HLHS tricuspid valve verification results.} (\textbf{A}) A 3D image-derived patient-specific tricuspid valve geometry is shown. The valve leaflet geometry and papillary muscle tips were identified from 3DE images of the tricuspid valve of an 11-year-old patient with HLHS in 3D Slicer. A total of 69 branch chords were projected from the papillary muscle tips to a chordae insertion zone defined from the free edge to mid-height of the tricuspid valve. A uniform transvalvular pressure of 97~mmHg was applied to the ventricular surface of the tricuspid valve. (\textbf{B}) The mean and standard deviation of the estimated $E$ and $\nu$ in the Neo-Hookean model were 526.84$\pm$47.88~$\text{kPa}$ and 0.317$\pm$0.021 at the final analysis step. (\textbf{C}) The mean and standard deviation of the estimated $c_0$, $c_1$, and $c_2$ in the Lee-Sacks model from 10 repeated tests were 165.57$\pm$4.01~$\text{kPa}$, 18.68$\pm$10.74~$\text{kPa}$, and 2.09$\pm$1.25, respectively. (\textbf{D}) The simulated systolic tricuspid valve produced excellent agreement when overlaid on top of the medical image (in light blue registration results, in light green for the Neo-Hookean model and light yellow for the Lee-Sacks model) The mean symmetric distance (MSD) between the reference segmentation model and FE-simulated systole model was 0.972~mm using Neo-Hookean material model and 0.969~mm for the Lee-Sacks material model.}\label{fig:example4:1}
\end{figure}

Finally, we applied our PINN model to a patient-specific tricuspid valve. The diastolic geometry of the tricuspid valve (Fig.~\ref{fig:example4:1}A) was segmented from the 3D TEE images of an 11-year-old patient with HLHS. 3DE images were chosen because they are the most common imaging modality used to evaluate valvular dysfunction and have relatively high temporal resolution to facilitate accurate image registration. The valve configuration at mid-diastolic frame was assumed stress-free as the transvalvular pressure on the leaflets at this point in the cardiac cycle is trivial~\cite{pham2017fem, liang2017, mao2017}. Due to the rapid leaflet motion that occurred in the transitioning phase, the strong-form dynamic momentum equation, $\mathbf{\sigma}_{ij,j}=\rho \frac{\delta^2 u_i}{\delta^2 t}$, was the governing PDE. Here, $\mathbf{\sigma}_{ij}$ denotes the Cauchy stress (we derived Cauchy stress from the first Piola-Kirchhoff stress to maintain consistent notation in our work), $\rho$ denotes material density, $u_i$ is displacement, and $t$ is time. Transvalvular pressure of 97~mmHg was applied to the ventricular surface of the tricuspid valve. The coefficients within the isotropic Neo-Hookean ($E$ and $\nu$) and Lee-Sacks material hyperelastic models ($c_0$, $c_1$, and $c_2$) were determined to characterize the elastic properties of the leaflets.

Forward FEA was sought to verify the estimated Neo-Hookean and Lee-Sacks elastic parameters. The annulus displacements derived from image registration were used to enforce annulus dynamic motion.  Due to the challenges of identifying the chordae tendineae from 3DE images, we followed a similar chordal projection scheme as described in Wu et al.~\cite{Wu2022_JBME} However, in the current work, the chordae tendineae were modeled as linear elastic trusses, as opposed to springs, with a cross-sectional area of 0.8 $\text{mm}^2$, an average Young's modulus of 20 MPa, and a Poisson's ratio of 0.3~\cite{Krishnamurthy2008}. The tips of the papillary muscles assumed a pinned boundary condition. The FE model of the tricuspid valve was discretized int 16,523 tetrahedral elements and analyzed through transient dynamic analysis. All FEA were run using an open-source FE modeling software: FEBio~\cite{Maas2012}. The resulting leaflet deformation from each material model was compared with the reference segmentation model to validate the accuracy of the estimated elastic parameters.

\subsubsection{Estimated results from PINNs}
In this example, the same network architecture as in the previous example was used. Similar to the previous example, we did not explicitly constrain the pressure condition. While the transvalvular pressure in patients with systemic atrioventricular valve diseases (as in the present study) is comparable to the systolic blood pressure measured during echocardiography, accurately measuring transvalvular pressure in patients with normal valve anatomy typically requires invasive procedures such as catheterization. Therefore, we leveraged the flexibility of PINNs to simultaneously learn displacement, stress, and mechanical properties using reference displacement data and governing PDEs to demonstrate the broader applicability of the current framework. The leaflet deformation and net stress resulting from the ventricular pressure and chordal reaction forces were output from the parallel FNNs. Since the dynamic momentum equation is applied in this example, the input incorporates a time component, represented as (x, y, z, t). In the Neo-Hookean PINN model, the loss weights were $w_\text{PDEs} = w_\text{M} = 10$ and $w_\text{D} = 1$. The learning rate decay was set to $10^{-3}$ with a decay rate of 0.45 every 15,000 iterations. In the Lee-Sacks PINN model, the loss weights were $w_\text{PDEs} = w_\text{M} = 10^{-3}$ and $w_\text{D} = 1$. The learning rate decay was set to $10^{-3}$ with a decay rate of 0.62 every 15,000 iterations. We trained the neural networks using the Adam optimizer for 250,000 iterations. The loss term associated with the reference displacement data was minimized using the mean symmetric distance (MSD) metric. The average run time in this example was 3.2 hours.

In the Neo-Hookean model, the mechanical properties $E$ and $\nu$ were constrained to a broad range of [0, 800]~$\text{kPa}$ and [0, 0.5], respectively. The mean and standard deviation values of $E$ and $\nu$ from 10 repeated tests converged 526.84$\pm$47.88~$\text{kPa}$ and 0.317$\pm$0.021 (Fig.~\ref{fig:example4:1}B). In the Lee-Sacks, the mechanical properties $c_0$, $c_1$, and $c_2$ were constrained to [0, 200]~$\text{kPa}$, [0, 200]~$\text{kPa}$, and [0, 20], respectively. The approximated mean and standard deviation values of $c_0$, $c_1$, and $c_2$ from 10 repeated tests were 165.57$\pm$4.01~$\text{kPa}$, 18.68$\pm$10.74~$\text{kPa}$, and 2.09$\pm$1.25 (Fig.~\ref{fig:example4:1}C).

\subsubsection{Finite element verification}
Valve closure was simulated using patient-specific parameters estimated by PINNs. Fig.~\ref{fig:example4:1}D shows the simulated systolic tricuspid valve superimposed on the medical image at four cross-sectional planes. As illustrated, we observed a high degree of overlap between the simulated tricuspid valve (Neo-Hookean in light green and Lee-Sacks in light yellow) and the white signals associated with valve leaflets on the 3D TEE images. The $95^{\text{th}}$ percentile Hausdorff distance (HD) resulting from the Neo-Hookean and Lee-Sacks models were 3.10~mm and 3.22~mm. The MSD between the registration and FE-simulated systolic models was 0.972~mm using Neo-Hookean material model and 0.969~mm for the Lee-Sacks material model. Both values fell within the range of inter-observer variability observed for manual segmentation reported in the literature~\cite{Herz2021}. These similarity metrics were computed from a Python script. 

\begin{figure}[!h]
\centering
\includegraphics[width=0.85\textwidth]{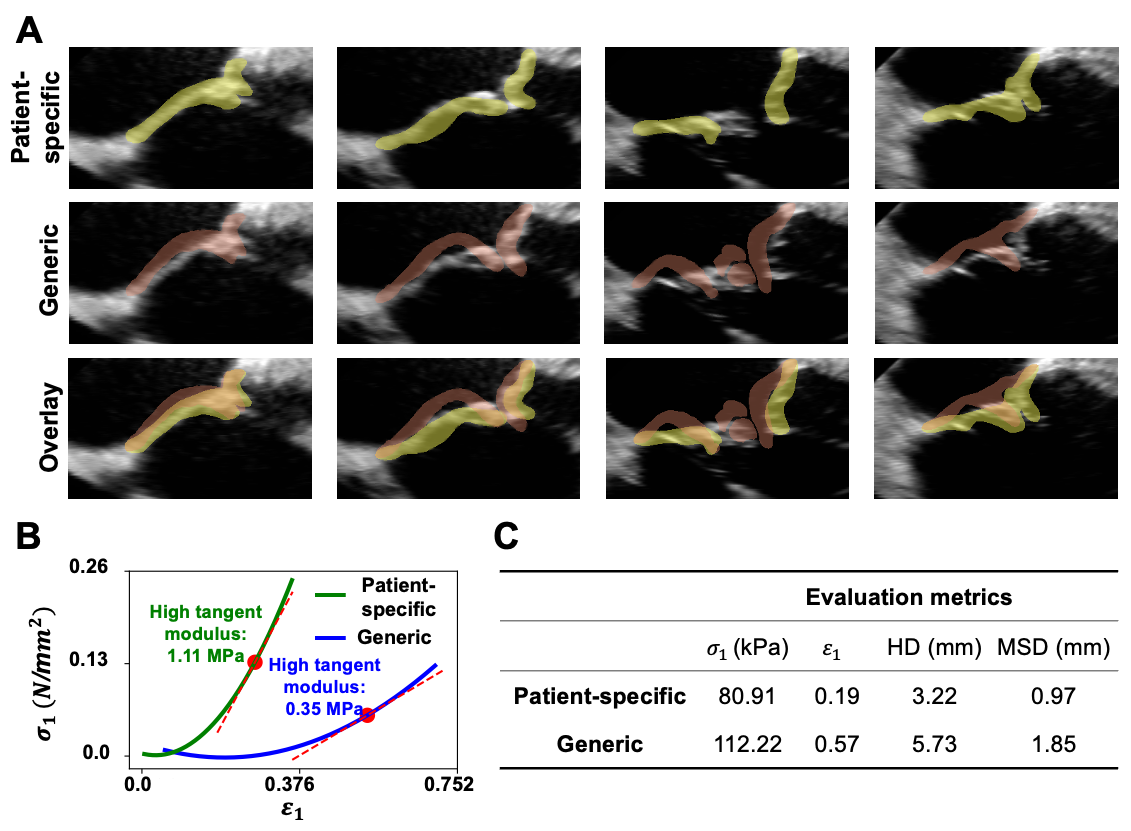}
\caption{\textbf{HLHS tricuspid valve using patient-specific vs generic tissue parameters.} The systolic geometries of the tricuspid valve model were generated in FEA with the Lee-Sacks constitutive law using patient-specific parameters derived from PINN and generic parameters~\cite{kamensky2018contact, johnson2021parameterization}. (\textbf{A}) Both models were superimposed onto the 3DE images to assess shape alignment. The FE model simulated with patient-specific parameters (top row) aligned more closely with the tricuspid valve signals in the 3DE images compared to the model using generic parameters (middle row). The generic parameters led to a consistent overprediction of leaflet displacements (bottom row). (\textbf{B}) The stress-strain curves for the two models are presented. The model using generic parameters exhibited a lower tangent modulus (reduced stiffness) compared to the patient-specific model. (\textbf{C}) The generic parameters resulted in higher values of $\sigma_{1}$ and $\epsilon_{1}$ than those produced by patient-specific parameters. Furthermore, the patient-specific parameters achieved a geometric alignment with the reference segmentation that was approximately twice as accurate as that obtained with the generic parameters.}\label{fig:example4:4}
\end{figure}

Fig. \ref{fig:example4:4} compares the mechanical responses and shape similarity metrics of the FE model with the Lee-Sacks constitutive law using patient-specific and generic elastic parameters~\cite{kamensky2018contact, johnson2021parameterization}. Fig. \ref{fig:example4:4}A illustrates the alignment between the simulated FE models and the reference 3DE images. The FE model with patient-specific parameters demonstrates better alignment with the tricuspid valve as captured in the 3DE images, while the model using generic parameters overestimates leaflet displacements. Fig. \ref{fig:example4:4}B presents the stress-strain curves for these models, derived by fitting a quadratic function to the nodal stress-strain data. The high tangent modulous, calculated as the slope of the curve at the 75$^{\text{th}}$ percentile of strain, of the model with generic parameters is more than three times lower than that with patient-specific parameters. This indicates that the generic parameters lead to a softer and more extensible mechanical response in the model. Fig. \ref{fig:example4:4}C compares the 95$^{\text{th}}$ percentile of the first principal stress, the first principal strain, and HD, as well as MSD, of the two models. In terms of geometric similarity, the generic parameters result in HD and MSD values of 5.73mm and 1.85mm, respectively, which are approximately twice as high as those generated using patient-specific parameters, indicating poorer alignment with reference segmentation.

\subsubsection{Uncertainty analysis of patient-specific tissue parameters estimated using PINN}
\begin{figure}
\centering
\vspace{-2ex}
\includegraphics[width=1\textwidth]{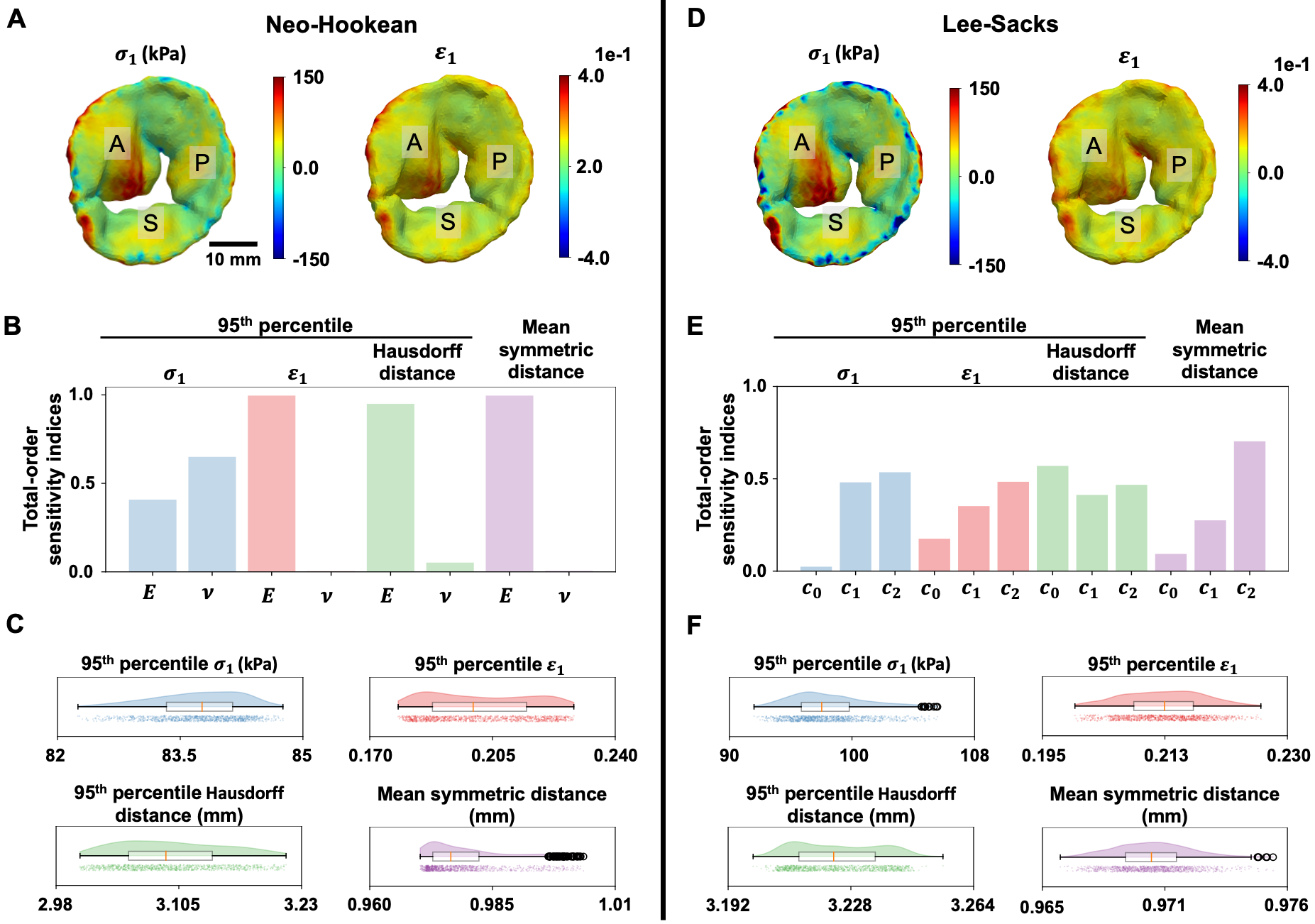}
\vspace{-2ex}
\caption{\textbf{HLHS tricuspid valve mechanical properties sensitivity analysis results.} The first principal stress ($\sigma_1$) and strain ($\epsilon_1$) simulated using the mean Neo-Hookean elastic parameters (\textbf{A}) and Lee-Sacks elastic parameters (\textbf{D}) are shown. A higher tensile stress and strain are observed on the anterior leaflet on both models. The 95$^\text{th}$ \%ile  $\sigma_1$ and $\epsilon_1$ resulting from the Neo-Hookean model are 75.02~kPa and 0.18 and from the Lee-Sacks model were 80.91~kPa and 0.19. In the Neo-Hookean model, (\textbf{B}) variations in $E$ have bigger effects on $\epsilon_1$, Hausdorff distance, and MSD, while variations in $\nu$ have a bigger effect on $\sigma_1$. (\textbf{C}) The percentage difference between the extrema is well above 10\% in $\epsilon_1$, indicating that strain prediction has the highest uncertainty with respect to changes in $E$ and $\nu$ among the metrics. The ranges between the extrema in other output metrics are relatively narrow. In the Lee-Sacks model, (\textbf{E}) $c_2$ has the highest influence on the 95$^\text{th}$ \%ile of $\sigma_1$, $\epsilon_1$, and the MSD, while $c_0$ has a higher sensitivity index at the 95$^\text{th}$ \%ile of Hausdorff distance. (\textbf{F}) $\sigma_1$ and $\epsilon_1$ demonstrate higher uncertainty, with more than 10\% difference between their extrema. However, the differences between the extrema are negligible in Hausdorff distance and MSD. Within each raincloud plot in (\textbf{C}) and (\textbf{F}), the left, center, and right vertical lines of the box plot indicate the 25$^\text{th}$ \%ile, median, and 75$^\text{th}$ \%ile of the data. The whiskers extend to the minima and maxima within 1.5 times the interquartile range of the box plot.}\label{fig:example4:2}
\vspace{-2ex}
\end{figure}

Here, we performed a sensitivity analysis to assess the effect of variations in each of the patient-specific elastic parameters on the 95$^{\text{th}}$ percentile (\%ile) first principal stress ($\sigma_1$), strain ($\epsilon_1$) and HD, as well as the MSD. Fig.~\ref{fig:example4:2}A-C are sensitivity analysis results generated for the Neo-Hookean model, and Fig.~\ref{fig:example4:2}D-F are the Lee-Sacks model. Fig.~\ref{fig:example4:2}A and D show the distribution of $\sigma_1$ and $\epsilon_1$ on the atrial surface of the tricuspid valve. These profiles were simulated using the mean Neo-Hookean elastic parameters (\textit{i.e.,} $E_\text{mean}$ and $\nu_\text{mean}$) and Lee-Sacks elastic parameters (\textit{i.e.,} $c_{0, \text{mean}}$, $c_{1, \text{mean}}$, $c_{2, \text{mean}}$), respectively. Qualitatively, both material models produced similar stress and strain patterns, with higher tensile stress and strain observed on the anterior leaflet compared to the other two leaflets. Quantitatively, the 95$^\text{th}$ \%ile $\sigma_1$ and $\epsilon_1$ resulting from the Neo-Hookean model were 75.02~kPa and 0.18 and from the Lee-Sacks model were 80.91~kPa and 0.19.

We used a Python subroutine called FEBioUncertainSCI~\cite{Burk2020, Bergquist2023} to calculate sensitivity of the 95$^\text{th}$ \%ile  $\sigma_1$, 95$^\text{th}$ \%ile  $\epsilon_1$, 95$^\text{th}$ \%ile Hausdorff distance, and MSD in the tricuspid valve with respect to variations in elastic parameters. Interested readers may refer to~\cite{Burk2020, Bergquist2023} on the theoretical background supporting the parametric uncertainty quantification method in UncertainSCI. In the Neo-Hookean example, we randomly generated 20 pairs of $E$ and $\nu$ in a 2-dimensional space bounded by $E_\text{mean}\pm E_\text{std}$ and $\nu_\text{mean}\pm \nu_\text{std}$ using a third order polynomial chaos expansion (PCE) function to build a PCE emulator. We performed a FEA for each pair of $E$ and $\nu$ to approximate their corresponding $\sigma_1$, $\epsilon_1$, Hausdorff distance, and MSD. These input and output variables were then used to train a PCE emulator to generate a manifold of total sensitivity indices between each permutation of the input and output parameters. The total sensitivity indices measure the combined direct effects of individual parameters (first-order sensitivity) and their fractional contributions resulting from higher-order parameter interactions (second or higher-order sensitivity) on the variance of model output parameters.

The total sensitivity indices at the 95$^\text{th}$ \%ile of $\sigma_1$, $\epsilon_1$, Hausdorff distance, and MSD in the Neo-Hookean model are provided in Fig.~\ref{fig:example4:2}B. As shown, within the sampling space, $E$ had significantly higher total-order sensitivity indices in  $\epsilon_1$, Hausdorff distance, and MSD, indicating its dominant role in influencing the kinematics behaviors of the tricuspid valve. In the context of $\sigma_1$, the sensitivity indices were comparable between $E$ and $\nu$, with $\nu$ had a slightly higher influence. We queried $10^3$ sampling pairs of $E$ and $\nu$ from the PCE emulator. Fig.~\ref{fig:example4:2}C presents raincloud plots that illustrate the key statistics (\textit{e.g.,} minimum, maximum, median) as well as the distribution of these $10^3$ $E$ and $\nu$ pairs on each output metric. Overall, we observed a narrow range of the maximum and minimum values in each output metric, except for the 95$^\text{th}$ \%ile $\epsilon_1$, where the percentage difference between the maximum and minimum values well exceeded 10\%. This indicates that strain prediction has the highest uncertainty with respect to changes in $E$ and $\nu$ among the metrics. 

In the Lee-Sacks example, we built a PCE emulator on 30 randomly sampled sets of $c_0$, $c_1$, and $c_2$ in a 3-dimensional space bounded by $c_{0, \text{mean}}\pm c_{0, \text{std}}$, $c_{1, \text{mean}}\pm c_{1, \text{std}}$, and $c_{2, \text{mean}}\pm c_{2, \text{std}}$ and their corresponding $\sigma_1$, $\epsilon_1$, Hausdorff distance, and MSD. The total sensitivity indices at the 95$^\text{th}$ \%ile  of $\sigma_1$, $\epsilon_1$, Hausdorff distance, and MSD are provided in Fig.~\ref{fig:example4:2}E. Overall, within this sampling space, $c_2$ appeared to have the highest influence on the 95$^\text{th}$ \%ile of $\sigma_1$, $\epsilon_1$, and the MSD. Contrarily, $c_0$ had a higher sensitivity index at the 95$^\text{th}$ \%ile  of Hausdorff distance. The distributions of each output metric from $10^3$ sampling pairs of $c_0$, $c_1$, and $c_2$ are shown in Fig.~\ref{fig:example4:2}F. Higher uncertainties in the 95$^\text{th}$ \%ile  of $\sigma_1$ and $\epsilon_1$ were observed, with more than 10\% difference between the maximum and minimum values of the metrics. However, the differences between the extrema were negligible in the 95$^\text{th}$ \%ile Hausdorff distance and MSD.

\begin{figure}[!ht]
\centering
\vspace{-2ex}
\includegraphics[width=1\textwidth]{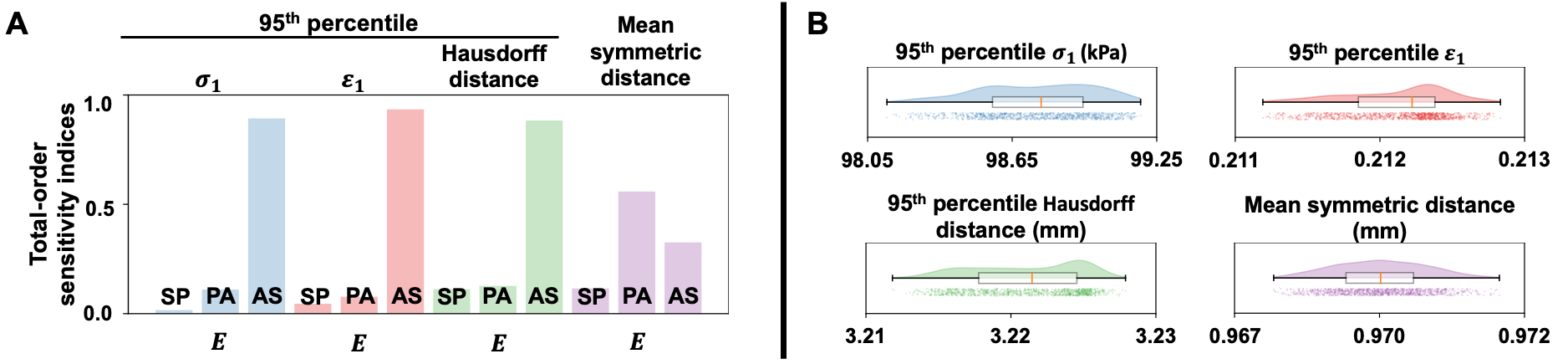}
\vspace{-2ex}
\caption{\textbf{HLHS tricuspid valve chordal sensitivity analysis results.} The effects of chorde mechanical properties on the 95$^\text{th}$ \%ile of $\sigma_1$, $\epsilon_1$, Hausdorff distance, and the MSD of the Lee-Sacks FE model were evaluated. (\textbf{A}) The 95$^\text{th}$ \%ile of $\sigma_1$, $\epsilon_1$, Hausdorff distance exhibited a high sensitivity to variations in the mechanical properties of the AS chordae, whereas MSD was most sensitive to alterations in the mechanical properties of the PA chordae. (\textbf{B}) The range between the maximum and minimum in each output metric was comparable, if not narrower, than those illustrated in Fig.~\ref{fig:example4:2}B. This observation indicates that variations in the mechanical properties of leaflets exert a greater influence on the 95$^\text{th}$ \%ile of $\sigma_1$, $\epsilon_1$, Hausdorff distance, and MSD of the FE model.}\label{fig:example4:3}
\vspace{-2ex}
\end{figure}

Notably, chordal structure and mechanical properties play a significant role in influencing proper valve closure. In our recent study~\cite{mangine2024chordaevariation}, we thoroughly investigated the uncertainty of chordal structure on functional metrics (\textit{i.e.,} strain and regurgitant orifice area). We found that variations in chordal structure had a minimal effect on leaflet strain and, on average, contributed to approximately 10 mm$^2$ difference in regurgitant orifice area. To provide a perspective on the effects of chordal mechanical properties on valve dynamics, we performed a sensitivity analysis to evaluate their impact on the 95$^\text{th}$ \%ile of $\sigma_1$, $\epsilon_1$, and Hausdorff distance, as well as MSD of the Lee-Sacks FE model. We varied the stiffness (E) of the septal-anterior (SP), posterior-anterior (PA), and anterior-septal (AS) chordae by $\pm 10\%$ of the reference value (E = 20 MPa). As demonstrated in Fig.~\ref{fig:example4:3}A, the mechanical properties of the anterior-septal chordae exhibited a dominant influence (having the highest total-order sensitivity indices) on the 95$^\text{th}$ \%ile of $\sigma_1$, $\epsilon_1$, Hausdorff distance. On the other hand, the posterior-anterior chordae had a leading effect on the MSD. A detailed statistic describing the distribution of each output evaluation metric from $10^3$ sampling pairs of $c_0$, $c_1$, and $c_2$ are presented in Fig.~\ref{fig:example4:3}B. The median values of the 95$^\text{th}$ \%ile of $\sigma_1$, $\epsilon_1$, Hausdorff distance, and MSD were 98.76 kPa, 0.21, 3.22 mm, 0.97 mm, respectively, as compared to the reference Lee-Sacks model values of 80.91 kPa, 0.19, 3.22 mm, and 0.97 mm. The close alignment between these values indicates that uncertainty in the mechanical properties of the chordae has a negligible impact on the evaluation metrics.

\section{Discussion}
\label{discussion}
We have developed a noninvasive approach to characterize the elastic properties of heart valve tissues using \textit{in vivo} 4D images. Our approach consists of two steps. First, deformable image registration was used to estimate the spatial displacements of valve leaflets from echocardiogram data. Subsequently, PINN was applied to infer the elastic parameters from image-derived displacement and the governing principles of the system. We evaluated the robustness of PINN on a series of simplified benchmarks prior to applying our framework to the challenge case, the tricuspid valve of a child. In the benchmark problems, PINN successfully determined the ``unknown" elastic parameters with satisfactory accuracy of within 3\% of $L^2$ relative errors compared to the ground truth. In the patient-specific tricuspid valve application, where the exact \textit{in vivo} material properties of valve tissue are unknown, a forward FEA was performed to simulate valve closure using the estimated tissue elastic parameters. Material parameters for the Neo-Hookean and Lee-Sacks constitutive models were identified. The resulting closed valve geometries were within 1~mm MSD compared to the segmentation ground truth. This error is within the range of those reported in the literature in the context of interobserver variability in manual valve modeling~\cite{Herz2021}. Furthermore, the patient-specific tissue parameters derived from our approach improved the accuracy of the simulated models by a factor of two compared to the generic parameters reported in the literature.

There has been a long-standing interest in \textit{in vivo} heart valve biomechanics research due to the strong implication and relevance of valve mechanics to the pathogenesis, progression, and repair of heart valves~\cite{Votta2013, Quarteroni2017, Bracamonte2022}. Numerous methods have been proposed to estimate \textit{in vivo} leaflet strain and valve dynamics to understand changes in strain between healthy and disease states, as well as between pre- and post-repair states~\cite{Aggarwal2016, Aly2021, Simonian2023, Laurence2024}. These retrospective studies are valuable for relating leaflet strain to valve function and identifying the functional characteristics of heart valves. Moving forward, it is equally imperative to develop the capability for prognostic studies to predict disease progression and to determine optimal repair options to reduce long-term patient morbidity and mortality~\cite{Sacks2019, Wu2023_JMBBM}. Our proposed approach provides a risk-free platform for clinicians to simulate outcomes of virtual intervention prior to performing surgery or intervention on actual patients~\cite{Kong2018, Aguilera2022, mathur2022}. Our framework, which we have named ADEPT, enables accurate identification of \textit{in vivo} elastic properties of valve tissue unique to each patient. This new capability offers the future potential to facilitate personalized virtual intervention analysis and use simulated models to identify the most effective repair option rather than clinical intuition alone. Due to the wide range of tissue extensibility from neonates to adults~\cite{Stephens2010, pham2017}, and the considerable heterogeneity in valve structure~\cite{Nam2023}, this approach holds particular significance for application to patients with congenital heart disease.

\subsection{Comparison with existing literature}
Traditionally, inverse FEA has been a common approach to characterize \textit{in vivo} elastic properties of heart valve tissues~\cite{rausch2013_BMM, Krishnamurthy2008, lee2014, lee2017vivo}. In these previous studies, physical markers were sutured to the leaflets to measure the actual leaflet displacements. Subsequently, FEA was performed iteratively to identify an optimal set of tissue stiffness parameters that provides agreeable simulated displacements compared to the measured ones. Of the studies, Rausch et al.~\cite{rausch2013_BMM} found elastic parameters of $c_0$ = 35.9~kPa, and weighting factors of $c_1$ = 2.3 and $c_2$ = 9.8 using a coupled anisotropic model provide the most agreeable in vivo behavior approximation of a mitral valve anterior leaflet. Krishnamurthy et al.~\cite{Krishnamurthy2008} performs inverse studies on 17 ovine anterior leaflets using a three-parameter orthotopic linear elastic model and found the elastic parameters to be $E_\text{circ} = 43 \pm 18$ kPa, $E_\text{rad} = 11 \pm 3$ kPa and $G_\text{circ-rad} = 121 \pm 22$ kPa. Lee et al.~\cite{lee2014} reported $c_0$ = 200~kPa, $c_1$ = 2968~kPa, and $c_2$ = 0.2661 for mitral valve leaflets within the isotropic Lee-Sacks model. Others have adopted $c_0$ = 10~kPa, $c_1$ = 0.209~kPa, and $c_2$ = 9.046 for modeling tricuspid valves~\cite{kamensky2018contact, johnson2021parameterization}. In our work, we furnished an noninvasive approach that combines deformable image registration and PINN to identify \textit{in vivo} tissue properties using clinically acquired 3D images. In our patient-specific analysis, the elastic parameter within an isotropic Neo-Hookean model was found to be $526.84 \pm 47.88$ kPa. Additionally, the elastic parameters within an isotropic Lee-Sacks model were found to be $c_0=165.57 \pm 4.01$ kPa, $c_1=18.68 \pm 10.74$ kPa, and $c_2=2.09 \pm 1.25$. Notably, the elastic parameters in different constitutive models exert varying degrees of influence on the concavity of the stress-strain curve. Furthermore, factors such as subject age, valve type, and the inherent differences between animal and human tissue further complicate direct comparisons. In this study, the higher predicted values compared to \cite{kamensky2018contact, johnson2021parameterization} suggest stiffer \textit{in vivo} tissue responses of the tricuspid valve. This observation is consistent with findings reported in the existing literature\cite{rausch2013_BMM, Krishnamurthy2008}.

In previous analyses of tricuspid valves, Stevanella et al.~\cite{Stevanella2010} reports a maximum $1^\text{st}$ principal stress of 430~kPa~\cite{Stevanella2010} and circumferential and radial strains of 0.13--0.16 and 0.25--0.30 on the anterior leaflet belly at 23~mm Hg transvalular pressure. Kong et al.~\cite{Kong2018} reports $1^\text{st}$ principal stress in the range of 24--91~kPa and $1^\text{st}$ principal strain in the range of 0.07 to 0.21 at mid-systole under the same peak transvalvular pressure. In our example, due to systemic right ventricle in the 11-year-old patient, the tricuspid valve was subjected to a  97~mm Hg transvalvular pressure, a pressure loading close to that experienced by a mitral valve. To offer additional references on mitral valve stress and strain, Wang et al.~\cite{Wang2013} reported 160~kPa maximum principal stress at 110~mm Hg peak systolic pressure. Lee et al.~\cite{lee2014} found a maximum radial and circumferential stresses of 509.5$\pm$38.4~kPa and 301.4$\pm$12.2~kPa on the anterior leaflet at 90~mm Hg peak transvalvular pressure. Furthermore, Rausche et al.~\cite{Rausch2011} found a maximum principal strains of 0.13$\pm$ 0.047 in ovine leaflets. Our \textit{in vivo} analysis produced a maximum $1^\text{st}$ principal stress and strain around 75.02~kPa and 0.18 using a Neo-Hookean model, and 80.91~kPa and 0.19 using a Lee-Sacks model, which are within the range of those reported by Kong et al.~\cite{Kong2018}. Qualitatively, our findings are in line with the literature as we also observed highest $1^\text{st}$ principal stress and strain located on the anterior leaflet. The exact stress and strain values are likely affected by a combination of factors, including valve morphology, transvalvular pressure, and the chosen material model. It is possible that the lower principal values found in our study are attributed to the smaller valve size in our patient compared to that in an adult. Furthermore, the annulus displacements were prescribed based on displacement data obtained from image registration. The potential noise in the displacement field due to image artifacts may influence the strain and stress near the annulus. Nonetheless, our strain values are in close alignment with existing literature reporting. 

\subsection{Limitations and areas for future work}
Notably, the combination of patient specific leaflet structure and material properties will be powerful, additional challenges to truly patient specific modeling remain including resolving leaflet chordal insertions. Current cine \textit{in vivo} imaging modalities can reliably visualize the papillary muscles (where chordae tendineae emanate from the ventricular myocardium), but not individual chordal insertions at the level of the valve leaflets themselves~\cite{Sacks2019}. However, previous work by Khaligi et al.~\cite{Khalighi2019} demonstrates that functional equivalent chordal models can be created. As such, while chordae insertion locations may influence the local stress and strain distribution, functional equivalence can yield accurate leaflet deformation, as demonstrated in our analysis. Unlike Khaligi et al., a more realistic branching chordal model is utilized in the present work. The effects of chordae density, length, and insertion sites on the resulting stress, strain, regurgitation orifice area, and leaflet contact areas of atrioventricular valves are areas for further investigation~\cite{mangine2024chordaevariation}. Finally, all analyses are fundamentally limited by the fidelity and faithful capture of leaflet geometry by the underlying image-modality, as well as the temporal resolution of the study.

In this proof-of-concept study, an isotropic homogeneous constitutive model was assumed to characterize tricuspid valve tissue as has been typical in modeling to date~\cite{mathur2022, mathur2024, Wu2022_JBME, pandya2023}. Although we observe a favorable deformation agreement in the simulated tricuspid valve relative to its motion in 3DE imaging, the inclusion of tissue anisotropy and a more representative material constitutive model may provide additional insights into the in vivo stress and strain patterns influencing the leaflets, as well as their implications for the growth and remodeling of the tissue. Nonetheless, obtaining leaflet fiber orientation, a critical component in anisotropic material models, from 3DE images is challenging due to limitations in image quality. Fortunately, prior studies have indicated that the degree of anisotropy in tricuspid valve tissue is relatively minimal in comparison to that observed in its left heart counterpart~\cite{pham2017}, and that the impact of tissue anisotropy on the resulting leaflet deformation is negligible~\cite{Wu2018}. Consequently, isotropic models have been widely adopted in numerical studies of tricuspid valves~\cite{mathur2022, mathur2024, Wu2022_JBME, pandya2023}. Further, the current PINN model can be easily modified, similar to that in our previous work~\cite{Wu2024_SM}, to account for heterogeneity in tissue elasticity observed in diseased valves due to calcification or exogenous material from prior surgical repair.

While the proposed workflow is applicable to all heart valves, it may require adjustments to the network hyperparameters when applied to different valve types or material constitutive models to achieve optimal convergence and solution accuracy. In this study, hyperparameters in PINNs were manually tuned. Future work could leverage optimization methods to automate this process~\cite{EscapilInchauspe2023}. Additionally, architectural improvements such as mixed-precision~\cite{hayford2024speeding} and forward-mode automatic differentiation~\cite{cobb2024} could be utilized to further improve computational efficiency and accuracy. To facilitate training in PINNs, the displacement fields of the tricuspid valves were obtained noninvasively using deformable image registration. Deformable image registration is a widely used method for tracking morphological displacements over time. However, the spatial-temporal resolution and noise in medical images can influence registration accuracy. This selected registration method was verified against manual tracing for the aortic root and aortic wall~\cite{Aggarwal2023, stoecker2022}, which has demonstrated a high degree of precision. Ongoing efforts are focused on developing more robust registration methodologies that integrate medical image data with simulated heart valve dynamics to further enhance accuracy and robustness in registration results for heart valves~\cite{Wu2024_MIDL}.

\section{Conclusions} 
\label{conclusion}
We presented a noninvasive approach to identify \textit{in vivo} elastic properties of valve tissue from 4D medical images using deformable image registration and PINNs. The estimated elastic parameters and consequent leaflet deformation, stresses, and strains are in excellent agreement with reference solutions and values reported in the literature. The proposed approach of noninvasively deriving mechanical insights from clinically accessible data holds particular translational values for evaluating longitudinal pathological changes and monitoring clinical disease progression that would be difficult to achieve with conventional methods. While our work directly benefits the development of patient-specific computer simulations of heart valve mechanics and repairs, the idea of combining image registration and PINNs to determine elastic properties of soft tissue noninvasively, whether for diagnostic or prognostic purposes, is broadly applicable to a wide spectrum of cardiovascular structures and beyond.

\section*{Acknowledgments}
This work was supported by the Cora Topolewksi Pediatric Valve Center at the Children's Hospital of Philadelphia, an Additional Ventures Expansion Award, the National Institutes of Health (NIGMS 2R01GM083925, NHLBI T32 HL007915, NHLBI K25 HL168235, NHLBI R01 HL153166, NHLBI R01 HL163202), the U.S. Department of Energy (Grants No.~DE-SC0025592 and No.~DE-SC0025593), and the National Science Foundation (Grant No.~DMS-2347833). The authors thank the CHOP Research Institute for generously providing the GPU resources necessary to support this project.

\section*{Data availability}
Displacement training data will be made available upon publication on GitHub at \url{https://github.com/lu-group/adept}.

\section*{Code availability}
The code for this study will be made available upon publication on GitHub at \url{https://github.com/lu-group/adept}. 

\bibliographystyle{elsarticle-num}
\bibliography{library}

\end{document}